# Giant Anomalous Hall Conductivity and Gilbert Damping in Room-temperature Ferromagnetic Half-Heusler Alloys PtMnBi

Hong-Xue Jiang[1], Jia-wan Li[1], Shi-Bo Zhao[1], Jie Wang[1], Yusheng Hou[1,*]

[1] Guangdong Provincial Key Laboratory of Magnetoelectric Physics and Devices, Center for Neutron Science and Technology, School of Physics, Sun Yat-Sen University, Guangzhou, 510275, China

**Abstract**

Half-Heusler alloys have emerged as promising candidates for novel spintronic applications due to their exceptional properties including the high Curie temperature ($T_C$) above room temperature and large anomalous Hall conductivity (AHC). In this work, we systematically study the magnetic and electronic properties of PtMnBi in α-, β-, and γ-phase using first-principles calculations and Monte Carlo simulations. The three phases are found to be ferromagnetic metals. In particular, the α-phase PtMnBi shows a high $T_C$ up to 802 K and a relatively large Gilbert damping of 0.085. Additionally, the γ-phase PtMnBi possesses a non-negligible AHC, reaching 203 $\Omega^{-1}cm^{-1}$ at the Fermi level. To evaluate its potential in nanoscale devices, we further investigate the α-phase PtMnBi thin films. The Gilbert dampings of α-phase PtMnBi thin films varies with film thickness and we attribute this variation to the distinct band structures at the high-symmetry point Γ, which arise from differences in film thickness. Moreover, the 1-layer (1L) α-phase thin film retains robust ferromagnetism ($T_C$ = 688 K) and shows enhanced Gilbert damping (0.14) and AHC (1116 $\Omega^{-1}cm^{-1}$) compared to the bulk. Intriguingly, under a 2% in-plane biaxial compressive strain, the Gilbert damping of 1L α-phase PtMnBi thin film increases to 0.17 and the AHC reaches 2386 $\Omega^{-1}cm^{-1}$. The coexistence of giant Gilbert damping and large AHC makes α-phase PtMnBi a compelling platform for practical spintronic applications, and highlights the potential of half-Heusler alloys in spintronic device design.

[*]Corresponding authors: houysh@mail.sysu.edu.cn

# 1 Introduction

The limitations of conventional electronic devices have driven the research focus toward the spintronics that harness both spin and charge degrees of freedom simultaneously for information processing [1]. Among diverse magnetic materials that hold promise for advancing the spintronics, the Heusler alloys including both full-Heusler and half-Heusler alloys have emerged as a captivating research focal point. The first Heusler alloy $Cu_2MnAl$ discovered by Heusler exhibits ferromagnetism, despite that its constituent elements themselves are non-magnetic [2,3]. This unique feature sparked significant interest in exploring the potentials of Heusler alloys further. Moreover, the properties of Heusler alloys can be finely engineered by tuning their elemental composition [4,5]. Given the vast number of possible elemental combinations, Heusler alloys can realize a rich and diverse array of properties, such as superconductivity [6-9], half-metallic ferromagnetism [10-13], and high Curie temperature [14,15]. These remarkable properties render them promising candidates for practical applications in the next generation of spintronic devices.

The temporal evolution of the magnetization $\boldsymbol{M}$ in a ferromagnetic (FM) material is captured by the Landau-Lifshitz-Gilbert equation [16-18]

$$\frac{d\boldsymbol{M}}{dt} = -\gamma \boldsymbol{M} \times \boldsymbol{H}_{eff} + \frac{\alpha}{M_S} \boldsymbol{M} \times \frac{d\boldsymbol{M}}{dt} \quad (1).$$

The right of the Landau-Lifshitz-Gilbert equation is composed of two terms, namely, the precession term (i.e., the first term) and the damping term (i.e., the second term). The precession term captures the rotational motion of the magnetization $\boldsymbol{M}$ around the effective magnetic field, $\boldsymbol{H}_{eff}$, while the damping term accounts for the gradual decay of this precession[18]. The dimensionless parameter $\alpha$ in Eq. (1) is the Gilbert damping parameter, which characterizes the rate of energy dissipation during the precession of the magnetization $\boldsymbol{M}$. The performance of a wide range of spintronic devices, including hard disk drives, magnetic random-access memories and magnetic sensors, is governed by the Gilbert damping $\alpha$[19].

To date, only low Gilbert damping parameters are found in Heusler alloys. Among them, $Co_2$-based full-Heusler alloys, such as $Co_2MnSi$, $Co_2MnAl$, and $Co_2MnGe$, have been confirmed to exhibit ultra-low Gilbert dampings, in the order of $10^{-4}$ [20-22]. Recently, experimental measurements have revealed that the Gilbert damping of the half-Heusler alloy NiMnSb thin film reaches an order of magnitude of $10^{-3}$ [23,24]. This implies that Mn-based half-Heusler alloys may have a relatively higher Gilbert damping. On the other hand, the synergy between spin-orbit coupling (SOC) and time-reversal symmetry breaking could engender a fascinating transport phenomenon termed the anomalous Hall effect (AHE)[25], which has been extensively investigated[26,27]. Singh *et al*. [28] have successfully measured the AHC of PtMnSb (with a $T_C$ of 560 K) which is as high as $2.2 \times 10^3$ $\Omega^{-1}$cm$^{-1}$ at 2 K. Due to the heavier atomic mass and stronger SOC of Bi compared to Sb, the SOC in PtMnBi is enhanced, implying the potential for a larger AHC. So, it is of importance to perform a systematic investigation into the Gilbert dampings and AHE in PtMnBi.

In this work, the magnetic and electronic properties of α-, β-, and γ-phase PtMnBi are systematically investigated using first-principles calculations. Our results indicate that the $T_C$ of α-phase PtMnBi reaches 802 K, which is significantly higher than room temperature. Among the three phases, γ-phase PtMnBi displays the highest AHC at the Fermi level, achieving a remarkable value of 203 $\Omega^{-1}$cm$^{-1}$. Notably, each phase of PtMnBi exhibits a large AHC exceeding 1000 $\Omega^{-1}$cm$^{-1}$ when the Fermi level is shifted to about 2 eV, suggesting that the huge AHC of PtMnBi can be realized experimentally by doping electrons. Besides, the α-phase PtMnBi exhibits a relatively large Gilbert damping, in the order of $10^{-2}$. We find that the Gilbert damping varies with the thickness of PtMnBi thin films, which is attributed to the modulation of the band structure at the high-symmetry point Γ by the film thickness. Remarkably, 1-layer (1L) PtMnBi thin film shows an impressively high $T_C$ of 688 K, a giant AHC (1116 $\Omega^{-1}$cm$^{-1}$) and a large Gilbert damping ($\alpha = 0.14$). Intriguingly, when a 2% biaxial compressive strain is applied to 1L α-phase PtMnBi, its spin polarization is enhanced to 70.62%. Meanwhile, the AHC at the Fermi level reaches an extraordinarily large value of 2386 $\Omega^{-1}$cm$^{-1}$, and the Gilbert damping attains a big value up to 0.17. The coexistence of high spin polarization, substantial AHC, and giant Gilbert damping in 1L PtMnBi thin film endows it with remarkable potential for applications. Our findings demonstrate the promising potential of half-Heusler alloy PtMnBi for versatile applications in spintronic devices.

## 2 Computational methods

The density functional theory (DFT) calculations are conducted using the Vienna *ab initio* simulation package (VASP) [29]. Projector-augmented wave pseudopotentials are utilized to describe the interactions between core and valence electrons [30,31]. We treat Mn-3$d$4$s$, Pt-5$d$6$s$ and Bi-6$s$6$p$ as valence electrons. For the depiction of exchange-correlation interactions, we employ the generalized gradient approximation formulated by Perdew, Burke and Ernzerhof [32]. We adopt an energy cutoff of 400 eV and a $k$-mesh grid of 21×21×21 centered at the Γ point. The three phases of PtMnBi are fully relaxed until the forces on each atom are less than 0.01 eV/Å. The energy convergence criterion is $10^{-6}$ eV. Based on magnetic interaction parameters derived from DFT calculations, the $T_C$s are obtained through Monte Carlo simulations.

Within the framework of scattering theory, the Gilbert damping parameters can be ascertained by first-principles calculations via the linear response formalism [33-35].

The reliability of this approach has been verified in previous studies [36-38]. By extending the torque method initially developed for studying magnetic anisotropy energy [39,40], the Gilbert damping can be calculated by employing the following formula [41]:

$$\alpha_{\mu\nu} = -\frac{\pi\hbar\gamma}{M_s}\sum_{ij}\langle\psi_i\left|\frac{\partial H}{\partial u_\mu}\right|\psi_j\rangle\langle\psi_j\left|\frac{\partial H}{\partial u_\nu}\right|\psi_i\rangle \times \delta(E_F - E_i)\,\delta(E_F - E_j) \quad (2).$$

In Eq. (2), $M_s$, $\gamma$ and $E_F$ represent saturation magnetization, gyromagnetic ratio and the Fermi level, respectively. $u_\mu$ is the deviation of a normalized magnetic moment away from its equilibrium. In Eq. (2), the delta function $\delta(E_F - E)$ is replaced by the Lorentzian function $L(E) = 0.5\Gamma/[\pi(E - E_0)^2 + \pi(0.5\Gamma)^2]$ with the scattering rate $\Gamma$ characterizing the temperature effect. To facilitate the calculations, we employ unit cells of three phases of PtMnBi to compute their Gilbert dampings, and simultaneously increase the $k$-point mesh to 31×31×31 to ensure the numerical convergence. When calculating the Gilbert dampings of α-phase PtMnBi thin films, a $k$-point mesh of 47×47×1 is used.

## 3 Results and Discussion

The half-Heusler alloy with a 1:1:1 stoichiometric ratio and general formula XYZ crystallizes in the face-centered cubic C1$_b$ crystal structure (space group F-43m, No. 216). The C1$_b$ crystal structure is constructed from three interpenetrating face-centered cubic sublattices. Constituting atoms occupy the Wyckoff positions 4a (0, 0, 0), 4b (1/2, 1/2, 1/2) and 4c (1/4, 1/4, 1/4), with the 4a-4b and 4a-4c atomic pairs forming NaCl-type and ZnS-type sublattices, respectively. Due to the presence of three nonequivalent element arrangements within this crystal structure, the half-Heusler alloy exhibits three distinct phases, namely, the α-, β- and γ-phase. For PtMnBi alloy, X and Y represent the transition metals Pt and Mn, respectively, and Z denotes the main group element Bi. Figures 1(a)-(f) illustrate the crystal structures of the α-, β- and γ-phase, in which Pt-Mn-Bi occupy the Wyckoff positions 4c-4b-4a, 4a-4b-4c and 4b-4c-4a, respectively.

Structural optimizations are performed using the unit cells of α-, β- and γ-phase PtMnBi. Their corresponding optimized lattice constants are 6.406, 6.465, and 6.310 Å, with their primitive cell energies of -19.412, -18.005 and -18.725 eV, respectively. It is evident that α-phase PtMnBi has the lowest energy, while γ-phase has the smallest lattice constant. These results are in good agreement with previous theoretical studies[42]. After determining the structural characteristics of the three phases of PtMnBi, we explore their magnetic properties. Among the three phases of PtMnBi, γ-phase exhibits the smallest magnetic moment (3.471 $\mu_B$/Mn), compared to 3.959 $\mu_B$/Mn in the α-phase and 3.949 $\mu_B$/Mn in the β-phase. These different magnetic moments can be qualitatively interpreted based on the structural configurations of the neighbor atoms. In the γ-phase, Mn atoms with Pt and Bi atoms as their nearest neighbors and possessing the lowest electronegativity among them, lose most of their electrons, thus giving rise to the smallest magnetic moment [43].

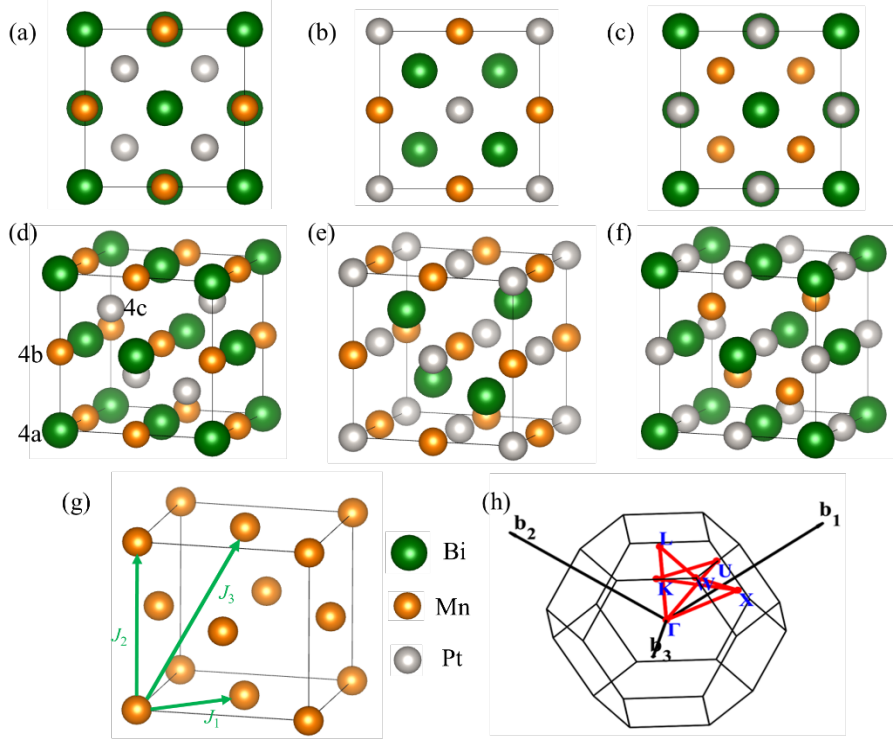

Fig. 1 Top and side views of crystal structures of (a) (d) α-phase, (b) (e) β-phase and (c) (f) γ-phase PtMnBi. (g) The first three nearest-neighbor Heisenberg exchange paths are shown by green arrows. (h) The first Brillouin zone of PtMnBi.

To determine the magnetic ground state of PtMnBi, a spin Hamiltonian composed of the first three nearest-neighbor (NN) Heisenberg exchange interactions is adopted. The spin Hamiltonian takes the following form:

$$H = \sum_{ij} J_{ij} \mathbf{S_i} \cdot \mathbf{S_j} \quad (3).$$

In Eq. (3), $\mathbf{S_i}$ and $\mathbf{S_j}$ correspond to the spin at magnetic sites $i$ and $j$, respectively, and $J_{ij}$ defines the coupling parameter mediating the Heisenberg exchange interaction between spins $\mathbf{S_i}$ and $\mathbf{S_j}$. Note that a negative $J_{ij}$ denotes a FM Heisenberg exchange coupling, and a positive $J_{ij}$ denotes an antiferromagnetic (AFM) one. We employ the energy mapping based on first-principles calculations to estimate Heisenberg exchange interaction parameters[44]. Table S1 and Fig. S1 in the supplementary material (SM) provide the constructed magnetic configurations and the magnetic moments of each Mn atom in α-phase PtMnBi, respectively. As illustrated in Fig. 2(a), our calculations reveal that the Heisenberg exchange parameters for all three phases of PtMnBi are negative, indicating that all of them could exhibit a FM ground state. Explicitly, the calculated $J_1$ for the α-, β- and γ-phase PtMnBi are -11.461, -0.037, and -0.497 meV, respectively. The corresponding $J_2$ are -5.276, -5.219, and -4.114 meV, and the $J_3$ are -2.753, -1.676, and -0.515 meV. We can see that the values of $J_1$, $J_2$ and $J_3$ of α-phase PtMnBi are the largest among the three phases, indicating that it possesses the strongest FM Heisenberg coupling interactions. In different phases, the variation of $J_1$ is remarkably pronounced, and the changes in $J_2$ and $J_3$ remain relatively minor. To determine the magnetic anisotropy, we calculate the energies of α-, β- and γ-phase PtMnBi with their magnetizations being along the [100], [011] and [111] directions, respectively. As shown in Fig. 2(b), the results suggest that the magnetic easy axes of α-, β- and γ-phase are [100], [011] and [111], respectively.

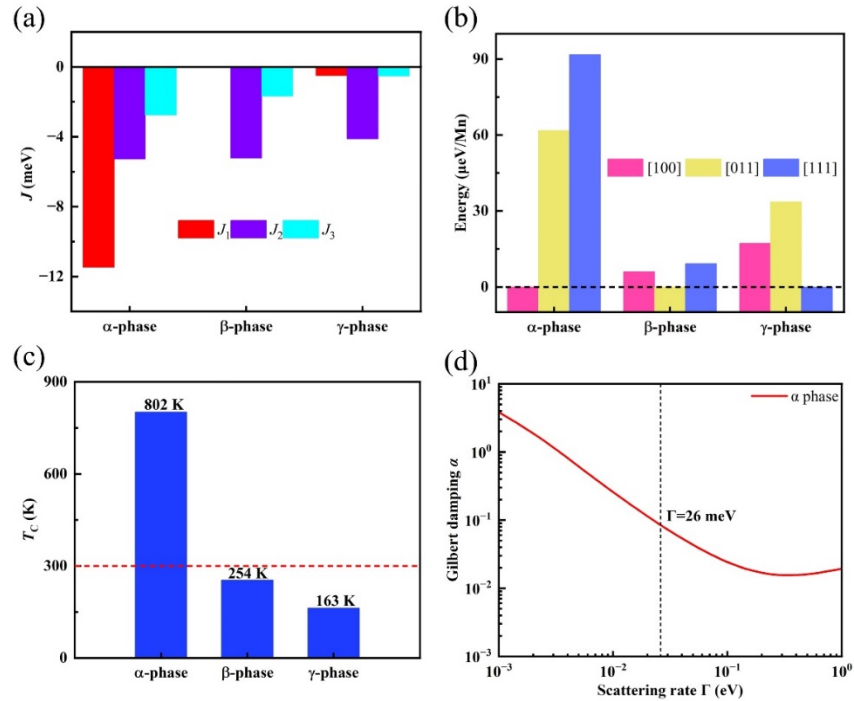

Fig. 2 Magnetic properties of PtMnBi. (a) Heisenberg exchange parameters $J$. (b) DFT calculated energies when the magnetization directions are along [100], [011], and [111]. For each phase, the lowest energy is set as the reference. (c) $T_C$s obtained by Monte Carlo simulations. (d) The Γ-dependent Gilbert dampings.

Within the mean field approximation, the $T_C$ of a ferromagnet can be estimated using the following relationship [44]:

$$T_C = \frac{S(S+1)}{3k_B} \sum_i z_i J_i \quad (4).$$

In Eq. (4), the summation is run over all the near neighbors of a given spin site, with $z_i$ representing the coordination number of adjacent neighbors connected by $J_i$. Based on Eq. (4), we see that the $T_C$ exhibits a direct proportionality to $z_i$ and the strength of the Heisenberg exchange interaction $J_i$. Using Eq. (4), the $T_C$s of PtMnBi in α-, β- and γ-phase are calculated to be 1389, 418, and 250 K, respectively. Given that the mean field approximation tends to overestimate magnetic transition temperatures [45], we also employ Monte Carlo simulations to determine the $T_C$ of PtMnBi. Excitingly, our results indicate that the $T_C$ of α-phase PtMnBi is as high as 802 K [Fig. 2(c)]. The $T_C$s of β- and γ-phase PtMnBi are 254 and 163 K, respectively. The high $T_C$, especially above the room temperature, is crucial for their potential applications in practical spintronic devices.

Figure 2(d) depicts the Γ-dependent Gilbert dampings of α-phase PtMnBi. We see that its Gilbert damping exhibits a characteristic behavior: it initially decreases and subsequently increases as the scattering rate Γ rises. This trend is in accordance with previous studies [19,41,46,47]. Within the framework of the breathing Fermi surface model, the Gilbert damping is composed of both intraband and interband contributions. As the scattering rate Γ increases, the interband contribution increases whereas the intraband contribution decreases, thereby resulting in a non-monotonic dependence of the Gilbert damping on the scattering rate [48]. Considering the practical applications of half-Heusler alloy PtMnBi, we focus on discussing its Gilbert damping at Γ = 26 meV, which corresponds to room temperature. At room temperature, the α-phase PtMnBi exhibiting a Gilbert damping parameter of 0.085, which is much larger than the Gilbert damping of body-centered cubic (bcc) Fe ($\alpha$ = 0.0013) [19]. The higher Gilbert damping $\alpha$ is, the more rapidly energy is dissipated, allowing the magnetization to reach equilibrium more quickly [49]. This rapid relaxation behavior is of vital importance to magnetic storage technologies.

Now, we investigate the electronic structures of different phases of PtMnBi. The electronic band structures of α-, β- and γ-phase PtMnBi are calculated along the high-symmetry paths X-W-L-Γ-X-K-U-Γ. Figures S2(a), (c) and (e) in the SM show their band structures when SOC is not included in our DFT calculations. These band structures indicate that three phases of PtMnBi are metallic. The projected density of states (PDOS) for the three phases given in Figs. S2 (b), (d) and (f) reveal that the electronic states near the Fermi level in all three phases of PtMnBi are predominantly contributed by Mn and Pt atoms. Considering the strong SOC of Bi atoms, we study the electronic band structures of PtMnBi with SOC [Figs. 3(a), 3(c) and 3(e)]. As shown in Figs. 3(b), 3(d) and 3(f), the PDOSs with SOC are similar to those without SOC. For α-phase PtMnBi, we see that SOC induces significant changes in the band structure near the Fermi level at the high-symmetry Γ point, characterized by the opening of a bandgap at crossing bands [Fig. 3 (a)]. By contrast, the β- and γ-phase PtMnBi exhibit relatively minor changes in their bands near the Fermi level at the Γ point under the influence of SOC.

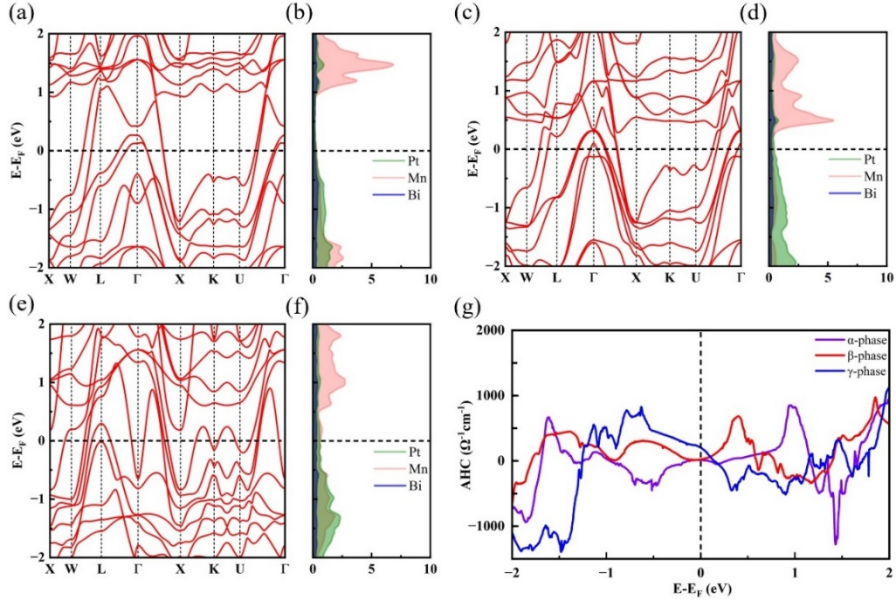

Fig. 3 Electronic properties and AHC of PtMnBi. (a) Band structures and (b) PDOS of the α-phase with SOC being considered. (c) and (d) are same as (a) and (b) but for the β-phase. (e) and (f) are same as (a) and (b) but for the γ-phase. (g) The dependence of AHC on the Fermi level position for the three phases of PtMnBi.

In ferromagnets where time-reversal symmetry is broken, the presence of strong SOC may give rise to AHE. To explore the AHE in PtMnBi, we construct an effective tight binding model via employing Wannier90 package [50]. The projected orbitals of Pt-$s, p, d$, Mn-$s, p, d$, and Bi-$s, p$ are used here. Figures S3(a)-2(c) demonstrate the close consistency between Wannier-interpolated band structures and DFT results. As shown in Fig. 3(g), the evolution of AHC with the Fermi level is calculated by WannierTools package [51] when the magnetizations of the three phases are aligned along their easy axes. It is noteworthy that the AHC of γ-phase PtMnBi reaches a large value of 203 $\Omega^{-1}\text{cm}^{-1}$ at the Fermi level. In contrast, the α- and β-phase PtMnBi exhibit significantly lower AHC, with values of 17 and 45 $\Omega^{-1}\text{cm}^{-1}$, respectively. Besides, within a range of 1.5 eV around the Fermi level, there are some very large values of AHC. Especially, at 1.43 eV above the Fermi level, the AHC of α-phase PtMnBi reaches a value of 1278 $\Omega^{-1}\text{cm}^{-1}$. For β-phase PtMnBi, AHC reaches a value of 686 $\Omega^{-1}\text{cm}^{-1}$ when the Fermi level is located at 0.40 eV. When the Fermi level is shifted to -1.48 eV, the AHC of γ-phase PtMnBi can reach a huge value of 1394 $\Omega^{-1}\text{cm}^{-1}$. These results suggest that the Fermi level can be tuned by doping electrons or holes to achieve larger AHC in experiments.

For practical applications, device miniaturization is becoming increasingly important. In recent years, thin films of the half-Heusler alloy NiMnSb have been successfully fabricated[52]. Inspired by this, studying the electronic structures and the Gilbert dampings of PtMnBi thin films is also of importance. Our interest is dedicated to α-phase PtMnBi, which stands out for its highest $T_C$ (802 K) and maximum Gilbert damping (0.085). Here, we construct different slab models with varying numbers of layers—specifically, 1L, 2-layer (2L), 3-layer (3L), 4-layer (4L), and 5-layer (5L)—to simulate thin films of α-phase PtMnBi. Each model is equipped with a vacuum space of 15 Å to prevent spurious interactions, and their structural configurations are shown in Fig. 4(a) (depicting only the 1L and 2L systems) and Fig. S4(a) (depicting the 3L, 4L and 5L systems). Based on Eq. (3), we first investigate the magnetic properties of 1L α-phase PtMnBi thin film. The calculated exchange parameters $J_1$, $J_2$, and $J_3$ are -14.335, -9.780 and -30.243 meV, respectively, indicating FM couplings for all three interactions. Our Monte Carlo simulations confirm a FM ground state with a $T_C$ of 688 K. Meanwhile, we calculate its magnetic anisotropy energy and find that its magnetic easy axis is out-of-plane (i.e., $z$ axis). Considering the typical suppression of $T_C$ with reduced thickness, the $T_C$s of α-phase PtMnBi thin films should still remain above room temperature [53].

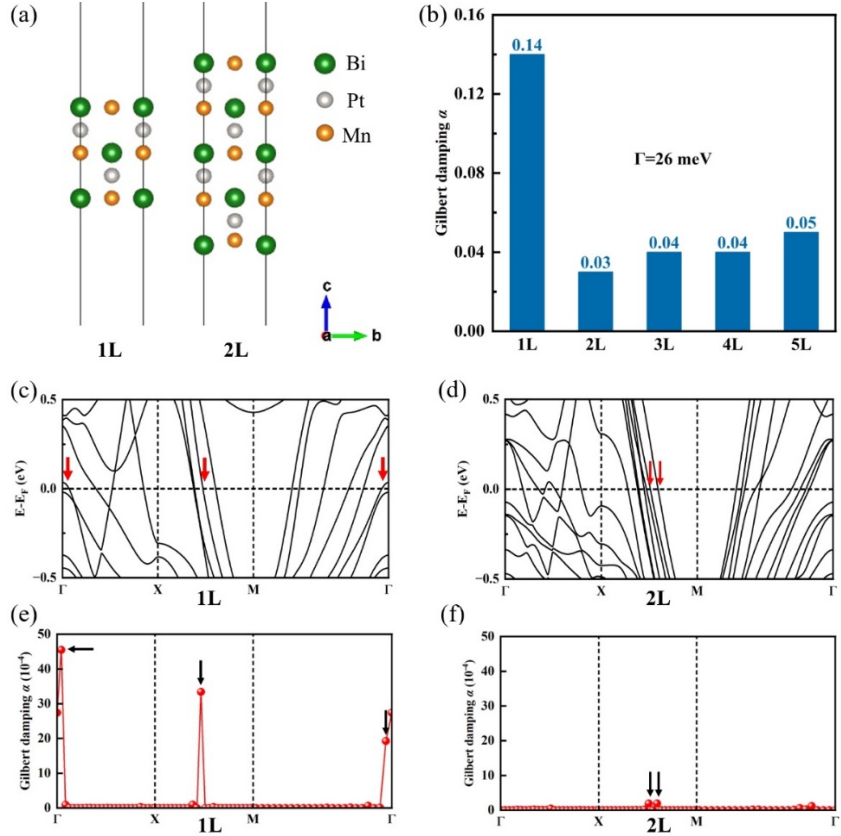

Fig. 4 Side views of crystal structures of (a) 1L and 2L α-phase PtMnBi thin films. (b) Thickness-dependent Gilbert dampings in α-phase PtMnBi films at Γ = 26 meV. (c) The band structure with SOC considered and (e) *k*-dependent contributions to the Gilbert damping of 1L α-phase PtMnBi thin film at Γ = 26 meV. (d) and (f) Same as (c) and (e) but for 2L α-phase PtMnBi thin film.

Given that α-phase PtMnBi thin films are room temperature ferromagnets, we next calculate their Gilbert dampings at the scattering rate Γ = 26 meV and illustrate the result in Fig. 4(b). One can observe that the Gilbert damping of 1L α-phase PtMnBi thin film is remarkably large up to 0.14, which significantly exceeds its bulk counterpart (0.085). When the thickness of the thin film exceeds 2L, the variation in Gilbert dampings becomes negligible. It is worth noting that, compared to 1L α-phase PtMnBi thin film, the Gilbert damping of 2L α-phase PtMnBi thin film is reduced by 78.6 %. To gain a deeper understanding of the variations in Gilbert dampings, we examine band structures and *k*-dependent contributions to Gilbert dampings in 1L and 2L α-phase PtMnBi thin films. As depicted in Figs. 4(c) and 4(d), compared with 1L α-phase PtMnBi thin film, the positions of bands crossing the Fermi level in 2L film undergo a significant shift. Especially, the bands at high-symmetry points, particularly those at the Γ point, exhibit distinct differences. By examining the *k*-dependent contributions to Gilbert damping [Figs. 4(e)-4(f)], we can see that the *k* points that make significant contributions are predominantly located in the proximity of the Fermi level (marked by arrows). It should be noted that not all *k* points with energies in the proximity of the Fermi level make significant contributions to Gilbert damping. Their contributions also depend on the matrix element $\langle\psi_i|\frac{\partial H}{\partial u_\mu}|\psi_j\rangle\langle\psi_j|\frac{\partial H}{\partial u_\nu}|\psi_i\rangle$. Moreover, in 1L α-phase PtMnBi thin film, the primary contributions to Gilbert damping originate from the vicinity of the high-symmetry Γ point and X-M path, whereas in 2L α-phase PtMnBi thin film it primarily stems from the X-M path only. In other words, the Gilbert damping of 2L α-phase PtMnBi thin film has sharply decreased because the bands near the high-symmetry Γ point have been tuned to a position far away from the Fermi level compared with those of 1L thin film. To gain an in-depth understanding of the mechanism for which the number of layers tunes the Gilbert damping, we also investigate the band structures and *k*-dependent contributions to Gilbert dampings in 3L, 4L and 5L α-phase PtMnBi thin films. As shown in Fig. S5 in the SM, the band structures near the high-symmetry Γ point also undergo significant changes, which are the main reason for the reduced contributions to the Gilbert dampings. Overall, by varying the thickness of α-phase PtMnBi thin films, one can effectively tune the band structures near the high-symmetry Γ point, thereby engineering the Gilbert damping.

During the experimental growth of half-Heusler alloy thin films, lattice mismatch induced by substrates can induce strain and changes in electronic band structures. Therefore, investigating the properties of α-phase PtMnBi thin films under in-plane biaxial strain is of significant importance for advancing their practical applications. Here, we focus specifically on the study of 1L α-phase PtMnBi thin film due to their remarkably large Gilbert damping.

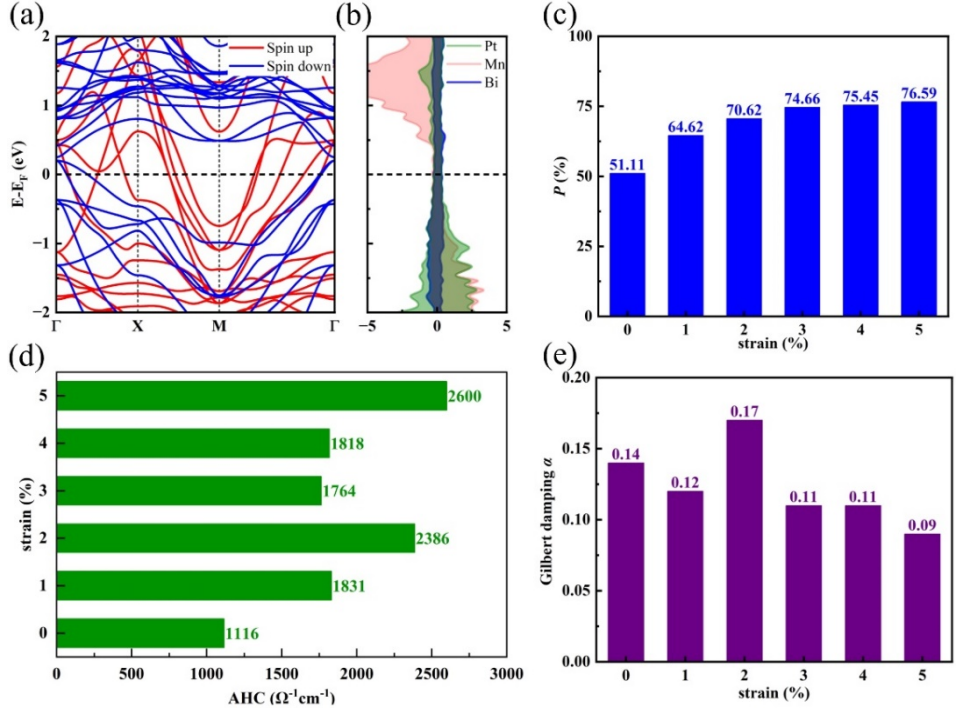

Fig. 5 Electronic and magnetic properties of 1L α-phase PtMnBi thin film. (a) Spin-polarized band structure and (b) PDOS. (c) The relationship between in-plane biaxial compressive strain and spin polarization, *P*. (d) The AHC values at the Fermi level under in-plane biaxial compressive strain when the magnetization direction is along the *z* axis. (e) The impact of in-plane compressive strain on Gilbert damping at $\Gamma = 26$ meV with the magnetization directions along the *z* axis.

The spin-polarized band structure and PDOS depicted in Figs. 5(a) and 5(b) demonstrate that 1L α-phase PtMnBi thin film exhibits metallic characteristics, with the electronic states near the Fermi level being primarily contributed by Pt, Mn, and Bi atoms. We observe that the bands crossing the Fermi level are predominantly provided by the majority spin states. In contrast, for the minority spin states, only the bands at the high-symmetry point Γ intersect the Fermi level. The spin polarization (*P*) can be calculated as follows [54],

$$P = \frac{N_\uparrow - N_\downarrow}{N_\uparrow + N_\downarrow} \quad (5).$$

In Eq. (5), $N_\uparrow$ and $N_\downarrow$ denote the total density of states (DOS) for the majority and minority states at the Fermi level, respectively. We approximate $P$ by employing the integral total DOS within 23 meV near the Fermi level. Figure 5(c) depicts the correlation between the in-plane biaxial compressive strain and $P$ in 1L α-phase PtMnBi thin film. Overall, the spin polarization of the system rises as the applied strain increases. Notably, when the applied stress reaches 2% or higher, the spin polarization exceeds 70%. This value is remarkably higher than that of the α-, β-, and γ-phase of bulk PtMnBi, which exhibit spin polarizations of 21.99%, 33.93%, and 33.23%, respectively. It is worth mentioning that when the lattice constant of α-phase PtMnBi is reduced from -3% to -11.2%, it will become a half-metal with 100% spin polarization[42].

To gain a comprehensive understanding of the AHE in 1L PtMnBi thin film, we calculate its AHC both before and after the application of biaxial compressive strain. As shown in Fig. 5(d), the AHC of 1L α-phase PtMnBi thin film at the Fermi level reaches an impressive value of 1116 $\Omega^{-1}\text{cm}^{-1}$, surpassing the AHC of all bulk phases of PtMnBi as well as that of bcc Fe (751 $\Omega^{-1}\text{cm}^{-1}$) [55]. Moreover, the application of compressive strain to thin film results in an increase in its AHC at the Fermi level. Notably, when the in-plane biaxial compressive strain is 2% and 5%, the AHCs at the Fermi level reach giant values of 2386 and 2600 $\Omega^{-1}\text{cm}^{-1}$, respectively. Figure S6 illustrates the distribution of Berry curvature in the first Brillouin zone of 1L α-phase PtMnBi thin film. We find that, regardless of whether compressive strain is applied, significant Berry curvatures exist near the point Γ, thus leading to the large AHC. In contrast, the application of tensile strain results in a reduction of the AHC at the Fermi level in 1L α-phase PtMnBi thin film (see Fig. S7(a) in the SM).

Figure 5(e) illustrates the strain-dependent variation in the Gilbert dampings of 1L α-phase PtMnBi thin film with out-of-plane magnetization orientation (i.e., $M \parallel z$ axis). Our calculations reveal that the Gilbert damping of 1L α-phase PtMnBi thin film remains larger than that of its bulk counterpart and other thicknesses of PtMnBi thin films, even after the application of in-plane biaxial compressive strain. Meanwhile, the application of compressive strain has a negligible effect on the Gilbert damping of 1L α-phase PtMnBi thin film. Specifically, as the external strain varies, the Gilbert dampings oscillate between 0.17 and 0.09. These indicate that the significantly large Gilbert dampings can be still maintained, despite the presence of lattice mismatch during the experimental fabrication of PtMnBi thin films. Besides, compressive strain effectively modifies the band structures near the Γ point in 1L system, thereby inducing changes in Gilbert damping. However, the large Gilbert damping originates from the contributions of all $k$-points within the first Brillouin zone (see Figs S8 and S9 in the SM). It is worth noting that tensile strain reduces the Gilbert damping of 1L α-phase PtMnBi thin film (see Fig. S7(b) in the SM). A higher Gilbert damping allows the magnetization to achieve equilibrium more swiftly, which is of paramount importance to magnetic storage technologies. Therefore, the 1L α-phase PtMnBi thin film, which features high spin polarization, a large AHC, and a significant Gilbert damping, holds tremendous potential in the design of advanced spintronic devices.

## 4 Conclusion

In summary, we systematically investigate the magnetic and electronic properties of α-, β-, and γ-phase PtMnBi through first-principles calculations and Monte Carlo simulations. Our results reveal that α-phase PtMnBi exhibits an exceptionally high $T_C$ of 802 K, with a room-temperature large Gilbert damping (0.085). We demonstrate that the band structures at the high-symmetry point Γ of α-phase PtMnBi thin films can be modulated by changing the film thickness, thereby enabling the control of their Gilbert dampings. Notably, the 1L α-phase PtMnBi thin film exhibits a remarkably high $T_C$ of 688 K, accompanied by a huge AHC of 1116 $\Omega^{-1}cm^{-1}$ and a giant Gilbert damping of 0.14. Interestingly, when a 2% in-plane biaxial compressive strain is applied, the spin polarization of 1L PtMnBi thin film reaches 70.62%, and the Gilbert damping is as high as 0.17. Moreover, the AHC (2386 $\Omega^{-1}cm^{-1}$) at the Fermi level is significantly enhanced. The coexistence of high spin polarization, large AHC and significant Gilbert damping in 1L PtMnBi thin film renders them highly promising for developing next-generation spintronic devices. Our work provides valuable theoretical guidance for the potential application of the half-Heusler alloy PtMnBi in spintronic devices.

**Declarations** The authors declare that they have no competing interests and there are no conflicts.

**Acknowledgements** This work was supported by the National Key Research Program of China (Grant No. 2024YFA1408303 and 2022YFA1403301), the National Natural Sciences Foundation of China (Grant No. 12474247 and 92165204). Yusheng Hou acknowledges support from Guangdong Provincial Key Laboratory of Magnetoelectric Physics and Devices (Grant No. 2022B1212010008) and Research Center for Magnetoelectric Physics of Guangdong Province (Grant No. 2024B0303390001). Density functional theory calculations are performed on Tianhe-Xingyi.